\documentclass[10pt, conference, letterpaper]{IEEEtran}

\IEEEoverridecommandlockouts

\makeatletter
\def\ps@headings{%
\def\@oddhead{\mbox{}\scriptsize\rightmark \hfil \thepage}%
\def\@evenhead{\scriptsize\thepage \hfil \leftmark\mbox{}}%
\def\@oddfoot{}%
\def\@evenfoot{}}
\makeatother 

\usepackage{times,amsfonts,amssymb,amsmath,setspace,color}
\usepackage{graphicx,url,bm,comment,subfigure,bbm}
\usepackage{epsfig}

\newcommand{\equaref}[1]{(\ref{eq:#1})}

\newcommand{\Uno}
{
\mathbb{I}
}

\newcommand{\diff}{{\rm\,d}}

\newcommand{\ls}[1]
   {\dimen0=\fontdimen6\the\font
    \lineskip=#1\dimen0
    \advance\lineskip.5\fontdimen5\the\font 
    \advance\lineskip-\dimen0
    \lineskiplimit=.9\lineskip
    \baselineskip=\lineskip
    \advance\baselineskip\dimen0
    \normallineskip\lineskip
    \normallineskiplimit\lineskiplimit
    \normalbaselineskip\baselineskip
    \ignorespaces
}

\newcommand{\tgifeps}[3]{
\begin{figure}[t]
\centering
\includegraphics[trim=0cm 1.0cm 0cm 0cm, width=#1cm]{#2.eps}
\caption{#3\label{fig:#2}}
\vspace*{-0.6cm}
\end{figure}
}

\newcommand{\sidebyside}[5]{
\begin{figure}
\begin{minipage}[t]{#5cm}
\includegraphics[trim=0cm 0.5cm 0cm 0cm, width=1.0\textwidth]{#1.eps}
\caption{#2\label{fig:#1}}
\end{minipage}
\hfill \hspace{0.2cm}
\begin{minipage}[t]{#5cm}
\includegraphics[trim=0cm 0.5cm 0cm 0cm, width=1.0\textwidth]{#3.eps}
\caption{#4\label{fig:#3}}
\end{minipage}
\hfill
\vspace*{-0.5cm}
\end{figure}
}

\newcommand{\sidebysidebyside}[7]{
\begin{figure*}
\begin{minipage}[t]{#7cm}
\includegraphics[trim=0cm 0.5cm 0cm 0cm, width=1.0\textwidth]{#1.eps}
\caption{#2\label{fig:#1}}
\end{minipage}
\hfill \hspace{0.2cm}
\begin{minipage}[t]{#7cm}
\includegraphics[trim=0cm 0.5cm 0cm 0cm,  width=1.0\textwidth]{#3.eps}
\caption{#4\label{fig:#3}}
\end{minipage}
\hfill \hspace{0.2cm}
\begin{minipage}[t]{#7cm}
\includegraphics[trim=0cm 0.5cm 0cm 0cm,  width=1.0\textwidth]{#5.eps}
\caption{#6\label{fig:#5}}
\end{minipage}
\hfill
\vspace*{-0.5cm}
\end{figure*}
}

\newcommand{\ba}{\begin{array}}
\newcommand{\ea}{\end{array}}
\newcommand{\be}{\begin{equation}}
\newcommand{\ee}{\end{equation}}

\newcommand{\short}{{\it  short }}
\newcommand{\llong}{{\it long }}

\begin{document}

\begin{sloppypar}

\title{
Efficient analysis of caching strategies under dynamic content popularity
}
\author{

Michele Garetto $\!^\dagger $,
Emilio Leonardi   $\!^\ast $,
Stefano Traverso  ${^{(*)}}$\\
$\!^\ast $ Dipartimento di Elettronica, Politecnico di Torino, Torino, Italy\\
$\!^\dagger $ Dipartimento di Informatica, Universit\`{a} di Torino, Torino, Italy
}


\maketitle
\begin{abstract} 
In this paper we develop a novel technique 
to analyze both isolated and interconnected caches 
operating under different caching strategies and realistic traffic conditions.
The main strength of our approach is the ability 
to consider dynamic contents which are constantly added into the system catalogue,
and whose popularity evolves over time according to desired profiles. 
We do so while preserving the simplicity and computational efficiency
of models developed under stationary popularity conditions, which 
are needed to analyze several caching strategies. 
Our main achievement is to show that the impact of
content popularity dynamics on cache performance can be effectively captured  
into an analytical model based on a fixed content catalogue (i.e., 
a catalogue whose size and objects' popularity do not change over time).
\end{abstract}

\maketitle



\section{Introduction}\label{sec:intro}

In the last few years the performance of caching systems has attracted
a renewed interest, especially in the networking community.
One reason for this revival can be attributed to the crucial role played
by caching in new content distribution systems emerging in the 
Internet. Thanks to an an impressive proliferation of cache servers, 
Content Delivery Networks (CDN) represent today the standard solution adopted 
by content providers to serve large populations of geographically spread users~\cite{Jiang_conext12}.
By caching contents close to users, we jointly reduce network traffic and 
improve user-perceived experience.

Another reason is the fundamental change of communication paradigm
that is gradually taking place in the Internet, from the
traditional host-to-host communication model to the new host-to-content 
paradigm. Indeed, a novel Information Centric Network (ICN) architecture has 
been proposed for the future Internet to better respond to the today and 
future (according to  predictions) traffic characteristics \cite{Jacobson-ICN}. 
In this architecture, caching becomes an ubiquitous functionality available at each router.

For these reasons it is of paramount importance to develop efficient tools
for the performance analysis of large-scale systems of interconnected caches 
for content distribution. Unfortunately, an exact analysis of
cache performance is notoriously a difficult task, considering that
the computational cost to exactly analyse just a single LRU 
(Least Recently Used) cache, grows exponentially with both the cache size 
and the number of contents \cite{King,Dan1990}.   

Many recent analytical efforts to evaluate the performance of both single
and interconnected caches leverage a simple yet powerful  
approximation technique known in the literature as Che's approximation, 
which was originally proposed in the seminal paper~\cite{che}.
This approximation, which has been recognized by many authors to be very accurate~\cite{Roberts1,Roberts_mix,amiciburini,nostro_infocom}, 
has opened the door to a flurry of new research efforts, which have extended 
the application of this approximation to a larger set of caching systems 
and traffic assumptions than those in which it was originally proposed.

In this paper, we put ourselves in the above research stream,
addressing one fundamental issue that still needs to be properly 
taken into account in the performance evaluation of caching systems, 
namely, the fact that contents to be cached can be extremely  
dynamic over time: new contents are steadily introduced in the set 
of available objects (think of YouTube), while their popularity
can exhibit a variety of patterns: for example, the popularity of 
some contents vanishes after a few days (e.g., sport news)
while others (e.g., songs or movies) attract requests for
prolonged time \cite{crane2008}.
In general, the number of requests attracted
by the contents can vary dramatically over time, and this can 
occur on time scales which are comparable to the churn time
of caches, making caching systems very challenging to analyze.

The effects of dynamic contents has only recently
being addressed in just a few studies (see Section \ref{sec:related}).
The large body of existing literature on cache systems simply ignores these effects,
assuming a stationary traffic model produced by a fixed catalogue of contents.
However, stationary traffic models are reasonable only  
when the cache churn time is small compared to the popularity
dynamics of contents. This assumption 
may no longer be considered acceptable in modern 
content distribution systems. 
Indeed, the increasing availability of inexpensive storage capacity
allows to store incredible amount of data in individual caches \cite{netflixopen}.
As consequence, the time-scale of cache dynamics becomes comparable or even larger than the lifetime 
of many objects, making unfeasible the assumption of constant object popularity. 
  
The main contribution of this paper is a novel technique to capture the impact
of dynamic contents on cache performance, while  preserving
the simplicity and accuracy of existing models 
based on the Che's approximation. In particular, our main achievement
is to show 
that it is possible to accurately capture the behavior of 
caching systems under dynamic content popularity (i.e., contents whose popularity evolves with  time) into a 
finite population analytical model (i.e. a model based on a fixed catalogue of contents), at the cost, however, of sacrificing 
one of the key properties of traditional models: the fact that request processes at 
different caches are independent.

Our modeling approach preserves many nice properties 
of stationary models (in particular, the possibility to analyze at low
computational cost many different caching strategies for both single
and interconnected caches), while allowing at the same time 
to consider the crucial role played by content popularity dynamics.

\section{System assumptions}\label{sec:prelim}

We start introducing some notation and assumptions.
In the simplest case, there is only one cache, whose size, expressed
in number of \lq objects', is denoted by $C$. 


The cache is fed by an exogenous arrival process of objects' requests
generated by users. Requests which find the object in the cache
are said to produce a \textit{hit}, whereas requests that do not find the object in 
the cache are said to produce a \textit{miss}. The main performance metric of interest
is the \textit{hit probability}, which is the fraction of requests producing a hit. 

In the case of cache networks, the \textit{miss stream} of a cache, i.e., the
process of requests which are not locally satisfied by the cache,
is forwarded to one or more caches (deterministically or at random), 
or to a common repository  storing entire object catalogue.
Eventually, all requests hit the target, and it is common in the modelling 
literature to neglect all propagation delays, 
including the delays necessary to 
possibly insert the object in one or more caches not storing it, in response to a miss. 

Cache systems and their analysis can be distinguished on the basis
of three main ingredients: i) the traffic model, i.e., the stochastic characterization 
of the request process generated by users; ii) the cache policy, i.e., 
how an individual cache reacts to a given object request; iii) the replication strategy,
i.e., how the entire cache network reacts to an object request, deciding in particular
in which caches objects get replicated back after a request hits the target.
We separately discuss each of the above ingredients in the next sections.

\subsection{Traffic models}\label{subsec:traffic}
We first recall the so-called Independent Reference Model (IRM), which
is de-facto the standard approach adopted in the literature to 
characterize the pattern of object requests arriving at a cache \cite{Coffman:73}. 
The IRM is based on the following fundamental assumptions: i) users request items
from a fixed catalogue of $M$ object; 
ii)  the process of requests of a given object is modeled by  a homogeneous Poisson process of intensity $\lambda_m = \Lambda p_m$.

The IRM is commonly used in combination with a Zipf-like law of probability $p_m$, 
which is the typical object popularity distribution observed in traffic measurements and widely adopted in
performance evaluation studies~\cite{zipf,Roberts_mix}.



By definition, the IRM completely ignores all temporal correlations 
in the sequence of requests. In particular, it does not take into account 
a key feature of real traffic usually referred to as {\em temporal locality},
i.e., the fact that, if an object is requested at a given point in time, 
then it is more likely that the same object will be requested again in the near future.
It is well known that traffic locality has a beneficial effect on cache performance
(i.e., it increases the hit probability)~\cite{Coffman:73} and several 
extensions of IRM have been proposed to incorporate it into a traffic model.
Existing approaches \cite{Coffman:73,virtamo98,amiciburini} 
typically assume that the request process for each object is 
stationary (i.e., either a renewal process or a Markov- or semi-Markov-modulated 
Poisson process).

One simple way to incorporate traffic locality in the traffic is the following.
Instead of a standard Poisson process (which produces an IRM sequence, as already said), 
the request process for a certain content at an ingress cache is described by an independent renewal process
with given inter-request time distribution. Let $F_R(m,t)$ be the cdf of the inter-request 
time $t$ for object $m$. The average request rate $\lambda_m$ for content $m$, which can be
expressed by $\lambda_m = 1/\int_0^\infty(1-F_R(m,t)) \diff t$, matches 
the desired average rate $\lambda_m = \Lambda p_m$. 
In the following, we will refer to the above traffic model as {\em renewal} traffic. 
As we will later see, these assumptions are not really 
appropriate to capture the kind of temporal locality usually
encountered in Video-on-Demand traffic,
because they cannot easily capture macroscopic, intrinsically
non-stationary effects related to content popularity dynamics.

Recently \cite{TraversoCCR} a new traffic model, named Shot Noise Model (SNM)
has been proposed as a viable alternative to traditional traffic models to capture macroscopic
effects related to content popularity dynamics. The  basic idea of the SNM is  to represent the overall request process as the superposition of many independent processes (shots), each referring to an
individual content. Specifically, the arrival process of requests for 
a given content $m$ at a cache is described by an inhomogeneous Poisson process of intensity 
$V_m h(t-t_m)$, where $V_m$ denotes the average number of requests attracted by the content,
$t_m$ is the time instant at which the content enters the
system (i.e., it becomes available to the users), and $h()$ 
is the (normalized) ``popularity profile'' of content $m$. 

SNM has been shown in~\cite{TraversoCCR} to provide a simple, flexible and accurate approach
to describing the temporal and geographical locality found in Video-on-Demand traffic.  
An interesting finding in~\cite{TraversoCCR} is that the particular shape of the
``popularity profile'' $h()$ has very little impact on the cache performance,
which essentially depends only on the average content life-span $L$.
This property actually plays a crucial role in our analytical methodology, as we will see.

To illustrate these facts, Figure~\ref{fig:phit_vs_cachesize_fertility_flip_trace_04} reports 
the cache size needed to achieve a desired hitting probability in a LRU cache fed by a real 
trace of YouTube video requests, which was kindly provided to us by the authors of~\cite{TraversoCCR}.
The trace was fitted by a multiclass SNM traffic model with 4 classes, 
all of them sharing the same shape for the ``popularity profile'' (but with different average life-span).
Results in Fig.~\ref{fig:phit_vs_cachesize_fertility_flip_trace_04} 
show that rather different shapes for the SNM (e.g., uniform vs power-law) produce
very similar curves, both in good agreement with results derived 
under the original Youtube trace. The curve labelled ON-OFF, also very close to 
the trace, can be obtained by adopting the methodology described in this paper, as explained later.
The plot contains also a curve labelled 'Naive IRM', corresponding to the cache performance 
observed after the application of a random permutation to the requests contained in the original trace: by so doing, 
the temporal locality present in the original trace is washed out, allowing us to assess 
the prediction error that one would get by following a naive IRM approach. 

\tgifeps{7.0}{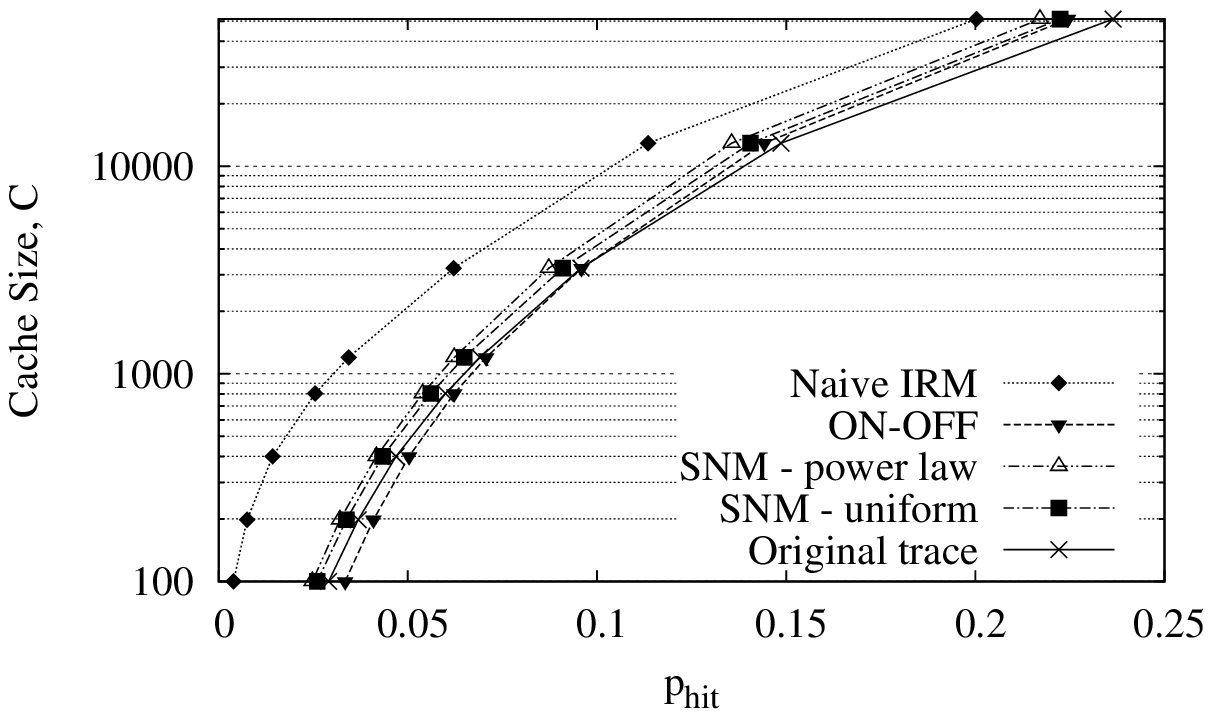}{Hit probability vs cache size, resulting from
feeding an LRU cache by: the original YouTube trace, a fitted multi-class SNM, our ON-OFF traffic model, and 
a reshuffled trace analogous to a naive application of the IRM model.}

\subsection{Cache policies} \label{subsec:policy}
In this work we will focus on the following strategies controlling the
behavior of an individual cache:
\begin{itemize}
\item {\bf LRU}: upon arrival of a request, an object not already stored in the cache is inserted into it.
If the cache is full, to make room for a new object the {\em Least Recently Used} item is evicted, i.e., the
object which has not been requested for the longest time.
\item {\bf q-LRU}: it differs from LRU for the insertion policy: upon arrival of a request, 
an object not already stored in the cache is inserted into it with probability $q$. 
\item{\bf RANDOM}:  it differs from LRU for the eviction policy: 
to make room for a new object, a random item stored in the cache is evicted.
\item{\bf 2-LRU}: this strategy, proposed in \cite{nostro_infocom}, is based on an 
effective, self-tuning insertion policy working as follows:
before arriving at the physical cache (storing actual objects),
requests have to traverse a virtual LRU cache put in front of it, 
which stores just object ID's. Only requests for objects whose ID is found in the 
virtual cache are forwarded to the physical cache.
The eviction policy at both caches, which for simplicity are assumed to be of the same size
(expressed either in terms of objects or ID's) is like LRU.  
\end{itemize}



\subsection{Replication strategies for cache networks}\label{subsec:rep}
In a system of interconnected caches, we need to specify
what happens along the route traversed by a request, after
the request eventually hits the target (in the worst case, ending up
at the repository containing all objects).  

We will consider the following mechanisms:
\begin{itemize}
\item {\bf leave-copy-everywhere (LCE)}: the object is 
put into all caches of the backward path.
\item {\bf leave-copy-probabilistically (LCP)}: the object is put with 
probability $q$ into each cache of the backward path.
\end{itemize}
An important property is the following: if we combine the LCP
replication strategy with standard LRU policy at all caches,
we obtain a cache system analogous to the one in which we adopt 
LCE replication in combination with q-LRU at all caches.
Hence, developing a model of q-LRU for individual caches
permits analysing LCP in a straightforward way.

We will not analyse in this work the leave-copy-down (LCD)
replication strategy, according to which the object is replicated only 
in the cache preceding the one in which it is found (if this is not an ingress cache). 
This would be an interesting direction of future research, in light of the 
excellent performance exhibited by this policy, which is
however more complex to analyze \cite{lao06}.

\section{Previous Work and Disussion}\label{sec:related}
Many recent efforts in modelling the performance of both isolated and interconnected caches 
leverage the Che's approximation originally proposed in \cite{che}, extending it along several directions. 
In \cite{Roberts1} authors provide a theoretical justification to Che's approximation,
showing that, asymptotically for large cache sizes, the cache eviction time $T_C$
satisfies a Central Limit principle. Papers \cite{Roberts1,nain,amiciburini,nostro_infocom} 
have extended Che's approximation to policies different from LRU, considering in particular 
RANDOM, FIFO, q-LRU, 2-LRU. The above caching policies have been analyzed in \cite{nain,amiciburini,nostro_infocom} 
also under more general traffic models than IRM, considering in particular the {\em renewal} traffic model 
introduced in \ref{subsec:traffic}, that allows capturing temporal locality in the traffic.
In all cases the application of Che's approximation provides a powerful
technique to decouple the behavior of different contents, essentially reducing cache dynamics to 
those of a simple single server queuing system under Poisson/renewal arrivals. 
All papers above, however, do not easily capture intrinsically non-stationary macroscopic effects 
related to content popularity dynamics.


As already mentioned, in \cite{TraversoCCR} authors have proposed a Shot Noise Model (SNM) to natively
describe the popularity evolution of new contents which are introduced into the catalogue.
Moreover, accurate analytical models still resorting on the Che's approximation can be developed 
for LRU caches (and networks) under SNM traffic.

Unfortunately, the SNM proposed in \cite{TraversoCCR} has some disadvantages. 
In particular, the analysis of non-LRU policies under SNM traffic turns out to be
very difficult. The reason for this is a bit technical, but it is worth explaining
it here so that the reader can better appreciate the contribution of our work.    
Under LRU, it is possible to write an explicit expression 
of the content $m$ hit-probability at time $t$ as $1- \Pr\{ \text{no requests for content $m$ arrive in $[t-T_C,t]$}\}$, 
which can be easily computed also under time-varying (inhomogeneous) Poisson processes.

However, under different caching policies such as RANDOM, q-LRU or 2-LRU,
an expression of the hit probability can be easily obtained only in the case of  
stationary (homogeneous) arrival process of content requests.
For example, under Che's approximation, dynamics of a RANDOM cache are reduced to those of a 
G/M/1/0 queue, being the content $m$  hit-probability  
equal to the probability of finding the server of this queueing system 
busy upon arrival. An explicit expression of this probability can be derived only under stationary 
conditions (i.e., at steady-state), whereas under non-stationary (transient) conditions 
the hit probability can only be expressed as a solution of a system of differential equations, 
making the computation excessively complicated. 

In this paper we propose a viable alternative to the SNM proposed in \cite{TraversoCCR}
to capture the impact of dynamic contents on cache performance, which allows
us to consider non-LRU policies at low computational complexity. 
We emphasize that, in the case of a single cache, our approach reduces to the application of existing
techniques developed for renewal traffic. However, in the case of cache networks,
our methodology departs completely form existing approaches, in that it assumes
request processes arriving at different caches to be strongly correlated,
in contrast to the standard independence assumption adopted in previous work.
 

\section{Modelling dynamic contents}\label{sec:approach}

We start describing our approach in the case of single cache. 
The basic idea is to capture the impact of dynamic contents (i.e., contents which start 
to be available in the system at a given point in time, and whose popularity evolves according 
to a certain profile), by using a stationary, ON-OFF traffic model associated to a properly chosen, 
fixed content catalogue of size $M$.

\tgifeps{8}{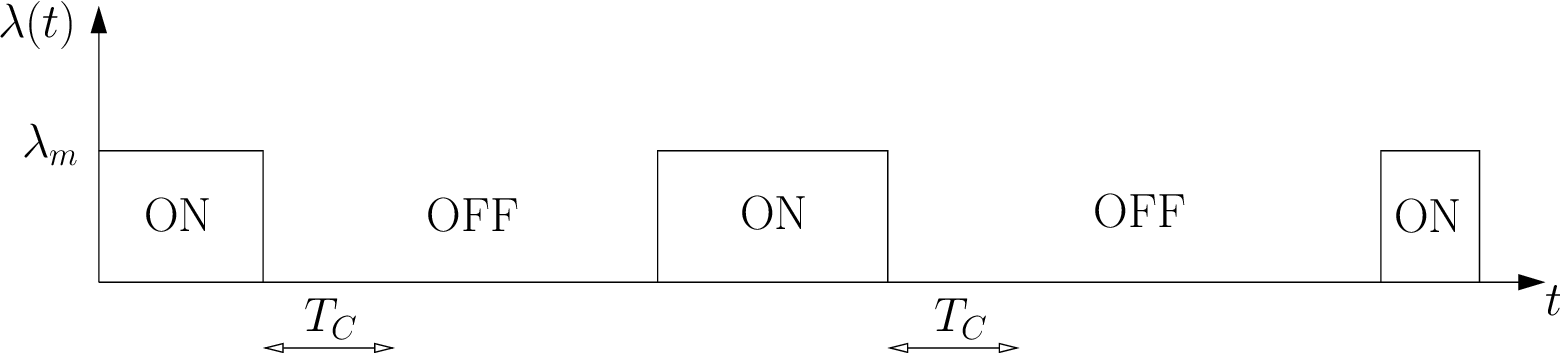}{ON-OFF modulated Poisson process describing the arrival of requests for a given content $m$.}
The rationale of our approach can be clarified with the help of 
Figure \ref{fig:approach}, which shows an ON-OFF modulated, homogeneous 
Poisson process describing the arrival process of requests for a given
content $m$ of our fixed catalogue. We assume that both ON and OFF periods are
exponentially distributed with mean duration $T_{\text ON}$ and $T_{\text OFF}$,
respectively. During an ON period, requests arrive with constant intensity
$\lambda_m$, which depends on the specific content $m$.
It follows that the average number of requests arriving during an ON period is  
given by: $V_m=\lambda_m T_{\text ON}$.

Suppose that $T_{\text OFF}$ is set much larger than the cache eviction time
$T_C$ ($T_{\text OFF}$ is a free parameter of our traffic model, hence it can   
always be set much larger than the maximum eviction time in the system). 
Then, at the end of the OFF period, the probability that the cache
stills contains a copy of object $m$ is negligible.  
Therefore, during the next ON period, content $m$ will produce an impact on the cache (in terms of hit probability) 
which is exactly the same as if it was a totally new content made available in 
the system at the beginning of the subsequent 
ON period. It follows that an ON period plays exactly the same role as a (rectangular)
shot in the SNM proposed in \cite{TraversoCCR}. 

Indeed, let us consider, for simplicity, a SNM in which all contents have the same temporal
profile, although they can attract a different average number of requests $V_m$ (heterogeneous objects in terms of 
popularity profile are handled by a multi-class approach, as done in \cite{TraversoCCR}. 
We exploit the observation made in~\cite{TraversoCCR} that the detailed shape of the popularity profile is 
not really important, while what really matters is its \lq effective duration' $L$ (called 
content life-span in \cite{TraversoCCR}). 
This means that we can well adopt a rectangular shape
for the ON period, whose duration $T_{\text ON} =  L$ is set equal to 
the first moment of the SNM profile.
Then, having chosen an arbitrarily large value of $T_{\text OFF} \gg T_C$,
we properly set the content catalogue $M$ so that the
average number of \lq active' contents is the same under both the SNM model
and the ON-OFF model. To do so, denoting by $\gamma$ the arrival rate of new contents
in the SNM model, we impose that
\be  \label{gamma}
 \gamma L = M \frac{T_{\text ON}}{T_{\text ON} + T_{\text OFF}}
\ee
from which we can derive the proper catalogue size $M$.
Note that the number of active contents is Poisson-distributed in the SNM model,
whereas it is binomially distributed under the ON-OFF model. However, it is
well known that the above two distributions are almost indistinguishable provided
that the mean number of active contents is large enough (say larger than a few tens), 
which is largely satisfied in all content distribution systems of interest, where
the number of available contents is in the order of thousands or millions.

At last, the values of $\lambda_m$ associated to contents of the fixed catalogue
are chosen so that the average number of requests produced during an ON period, which is
$V_m = \lambda_m \cdot T_{\text ON}$, has the same distribution as the number of requests 
produced by the shots in the SNM. 
Again, the catalogue size is usually large enough that we can consider the system ergodic, 
even if $\lambda_m$ remains the same for all ON periods associated to content $m$.  




As a proof of concept, we derived an equivalent ON-OFF traffic model for each of the 
four SNM classes in the experiment of Figure~\ref{fig:phit_vs_cachesize_fertility_flip_trace_04},
using the parameters reported in \cite{TraversoCCR}. Even in this complex scenario,
we observe a good agreement between the fitted SNM and the equivalent 
ON-OFF traffic model.
   
In the next Section we will show that our ON-OFF modulated Poisson traffic 
can be described by a standard {\em renewal} traffic model, which permits reusing
existing techniques to modeling the performance of various caching policies.

However, in our discussion so far we have considered just the simple case of one cache.
We still need to specify how to model the arrival processes of requests
arriving at the different ingress points of a cache networks. This raises
a subtle important point that marks a fundamental difference between our approach
and existing models in the literature.  

Previous models of cache networks under renewal traffic ~\cite{nain,amiciburini,nostro_infocom}  
assume that request processes at different ingress caches are independent.
We argue that this assumption is not appropriate in our case, because it would make ON periods
related to the same object of the catalogue totally uncorrelated from one ingress point to another,
washing out most of the temporal locality produced by  content popularity dynamics
that we are trying to capture in our model.

We therefore adopt exactly the opposite assumption, considering ON periods
associated to the same object to be {\em perfectly } synchronized among all ingress points.
This is reasonable, since new objects usually start to be 
available in the entire system at the same time. This means that
there exists a unique ON-OFF process for each object of the catalogue,
whose generated requests are split independently at random among the ingress caches
of the system (in proportion to the traffic volume arriving at each ingress cache).

To show the dramatic difference in cache performance obtained under the
above two assumptions (i.e., independent vs perfectly synchronized ON periods),
Fig. \ref{fig:confronto} reports the global hit probability
in a network of LRU caches having a binary-tree topology with four layers (15 caches).
The ON-OFF traffic is characterized by catalogue size $M = 3.5 \cdot 10^6$, $T_{ON} = 7$,
$T_{OFF} = 63$, while $V_m$ is Pareto-distributed with mean $10$ (at each ingress cache),
and scale-exponent $\beta = 2.5$.
The hit probability under the assumption of synchronized ON periods is about 4-times larger
than under the assumption of independent ON periods!
    
We conclude that our model based on synchronized ON-OFF processes is
dramatically different from existing models based on independent 
renewal traffic at the ingress caches.

\section{Modeling ON-OFF traffic as a standard renewal process}\label{sec:details}
We now show how previously defined ON-OFF process generates, for a given content of the
fixed catalogue, a sequence of requests which can be equivalently described by a standard renewal model.

Under the assumption that ON times are exponential distributed with mean $T_{\text ON}$, 
the number of requests  generated during an ON period turns out to be
geometrically distributed with parameter $p=\lambda_m/(\lambda_m+1/T_{\text ON})$ (starting from zero)
and average $V_m=\lambda_m T_{\text ON}$.

Indeed, by construction, the arrival process of requests follows patterns in which  
geometrically distributed sequences of \short inter-request times (with parameter $p$), 
taking place during  ON periods, are interleaved by sequences of \llong inter-request times
(again geometrically distributed with parameter $1-p$) occurring when the modulating process visits the 
OFF state. Figure \ref{fig:approach_new} illustrates the possible cases that can occur
in the generated sequence of requests. Note that when no requests are generated during an ON period
we get a combined longer inter-request time. When just one request is generated during an ON period,
two \llong inter-request times occur in sequence.     
 
\tgifeps{8}{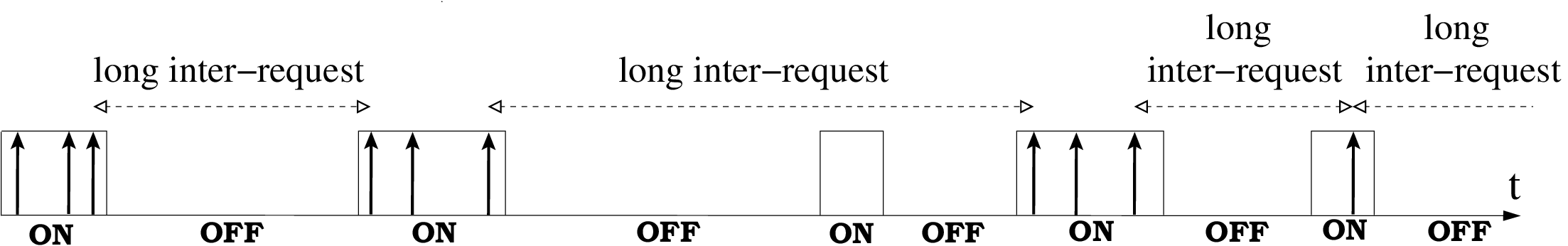}{Illustration of possible cases of inter-request time
for a given content produced by the ON-OFF model.}

Observe that \short inter-request times are exponentially distributed with parameter $\lambda_m+1/T_{\text ON}$.
An exact computation of \llong inter-request times is more involved, since it requires to evaluate the 
distribution of the interval between the last request occurring during an ON period and the 
next time at which a request is generated (which may incorporate ON periods in which no requests are generated) (see Figure
\ref{fig:approach_new}). 
Under the additional assumption that also OFF periods are exponentially distributed with mean $T_{\text OFF}$, 
an exact characterization of \llong inter-request times can be carried out by exploiting standard moment generating function
techniques (in this case \llong inter-request times are phase-type distributed).
However, this effort turns out to be unnecessary for our purposes, since, as long as the mean duration of 
the OFF period is much larger than $T_C$, the detailed shape of the distribution of \llong  
inter-request times has essentially no impact on cache performance. 
For this reason, we approximate \llong inter-request times by an exponential distribution 
matching only the first moment of the actual distribution of \llong inter-request times. 

To describe the process of requests arriving at non-ingress caches (in tree-like networks, caches
which are not leaves of the tree), we first need to characterize the miss stream
going out of previous caches. 
To do so, we adapted to our context techniques already presented in \cite{nain,amiciburini}.
As shown in~ \cite{nain}, under Che's approximation the miss stream of a cache fed by renewal traffic 
is again a renewal process. Indeed, the inter-miss distribution can be exactly characterized 
for a large class of cache policies, employing standard cycle-analysis of renewal processes. 

In our case, we describe the miss stream of a cache as an ON-OFF process 
having the same values of $T_{\text ON}$ and $T_{\text OFF}$ as the input process.
By so doing we can characterize again the miss stream as a renewal process whose inter-arrival 
times are partitioned into two classes of \short and \llong inter-miss times, 
inheriting the same semantic as before. 

\sidebysidebyside{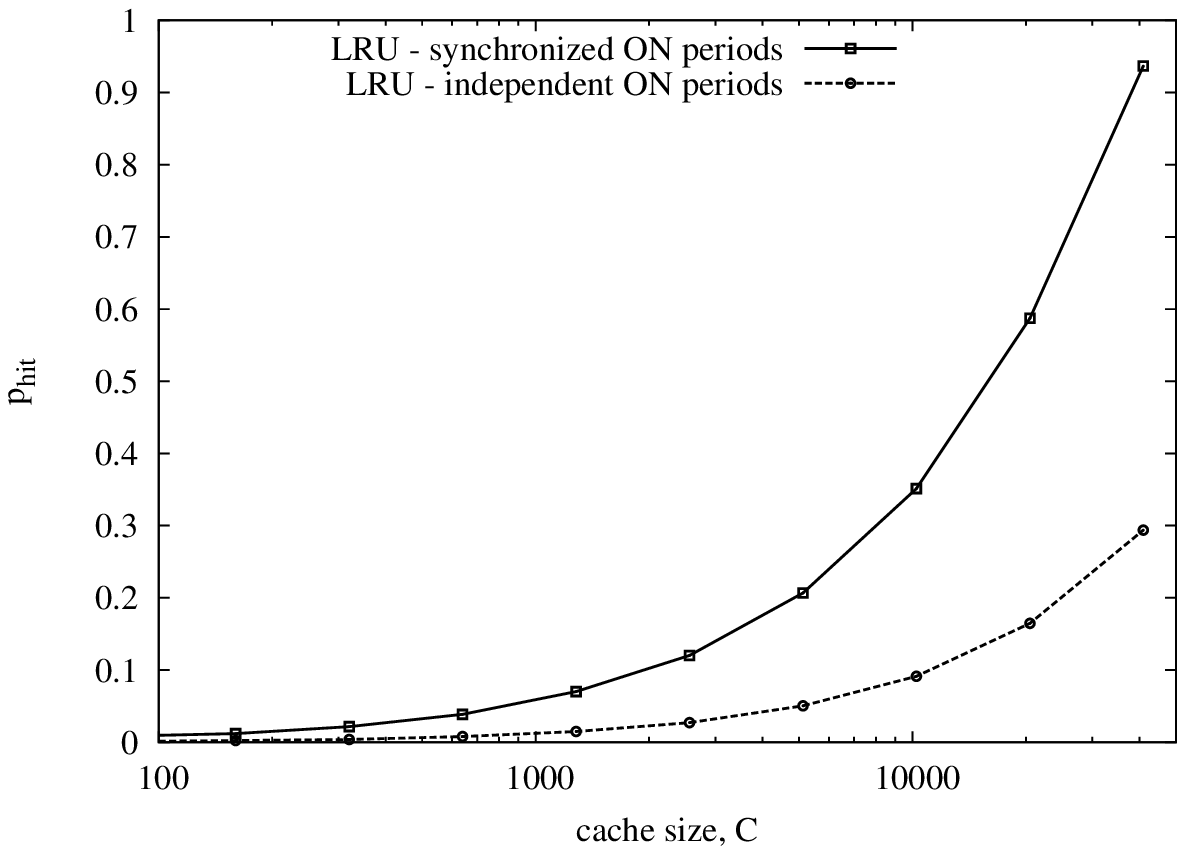}{Hit probability vs cache size, under different assumptions about the ON periods at the ingress caches of a tree network.}{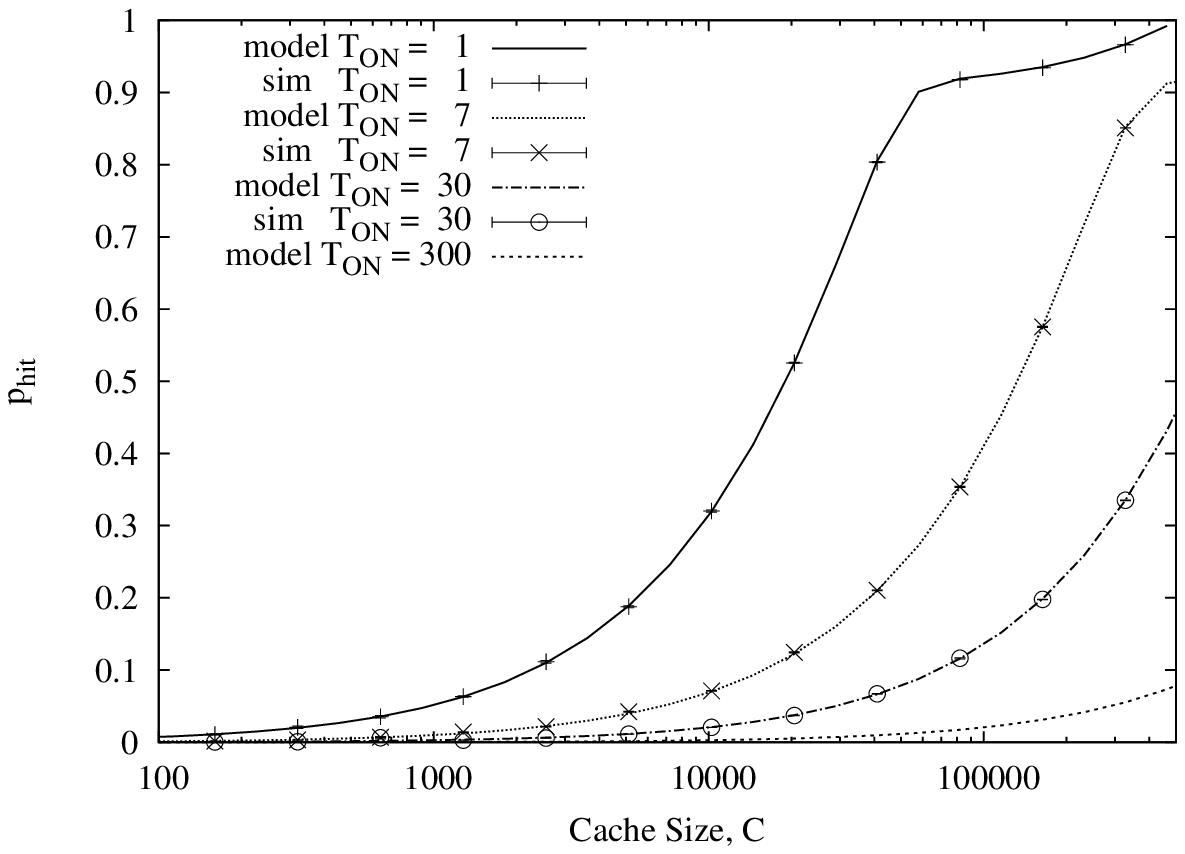}{Hit probability vs cache size, for different values of $T_{\text ON}$, under LRU.}
{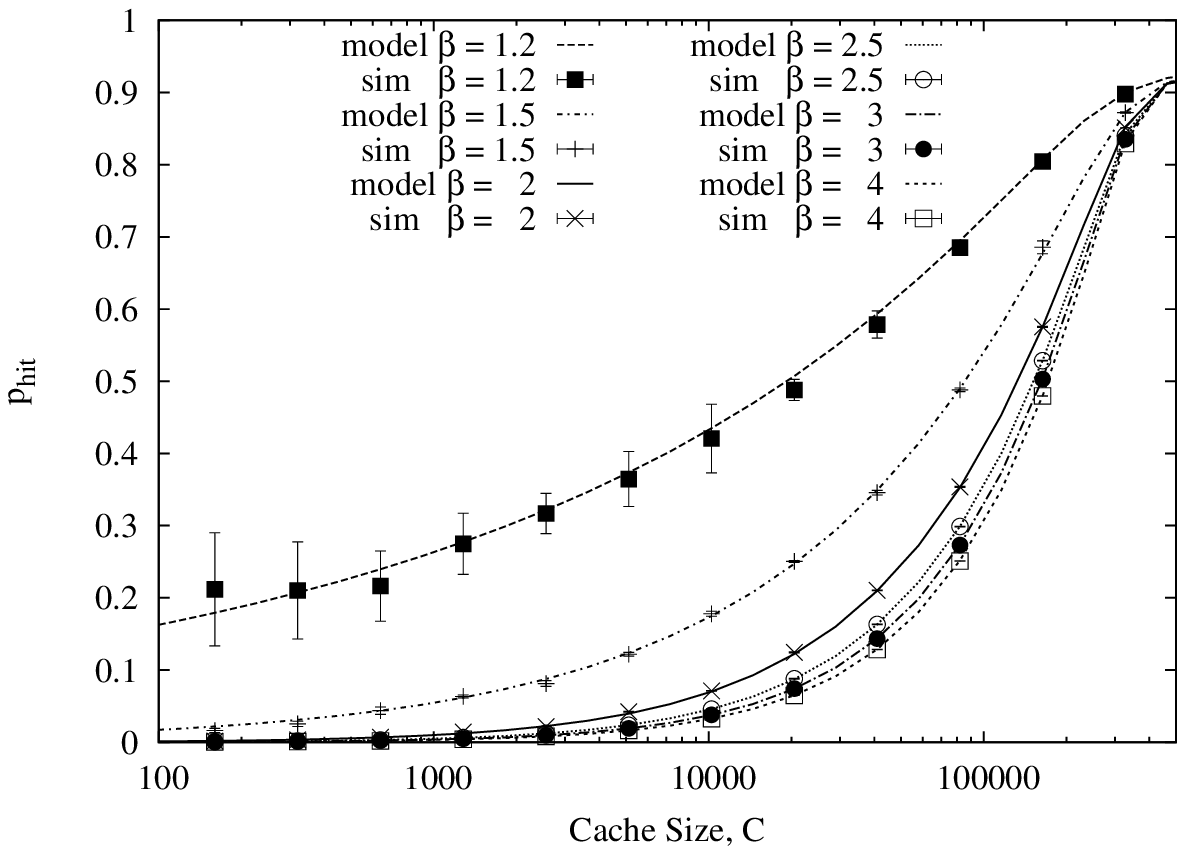}{Hit probability vs cache size, for different values of Pareto exponent $\beta$, under LRU.}{5.5}
 
In particular, \short inter-miss times (i.e., inter-miss times conditioned to the fact that the process keeps in 
ON) can be in principle exactly characterized following the approach in~\cite{nain}.  
In our model, however, to limit the computational complexity of the numerical solution, 
we prefer to adopt a second-order approximation, by selecting a priori a class of inter-miss distributions 
having two free parameters, which are set so as to match the first two moments of 
the exact {\em short} inter-miss time distribution.  

For LRU and RANDOM we consider the class of distributions given by a {\em shifted exponential}, i.e.,
\begin{equation}\label{eq:class1}
 F_{\text{short}}(m,t)=\left\{\begin{array}{cc} 1 & t\le T_m \\
                        e^{-\gamma_m (t-T_m)} &  t>T_m
 \end{array}\right.
\end{equation} 
For q-LRU we instead adopt a mixture of an exponential distribution (with weight $q$, and keeping the same 
parameter $\lambda_m$ of the inter-request distribution) and a shifted exponential distribution (with weight $1-q$), i.e.,
\begin{equation}\label{eq:class2}
 F_{\text{short}}(m,t)=\left\{\begin{array}{cc} (1-q)+q e^{-\lambda_m (t)}   & t\le T_m \\
                     q e^{-\lambda_m (t)}+(1-q)  e^{-\gamma_m (t-T_m)} & t>T_m \end{array} \right.
\end{equation}
Observe that in both classes above $\gamma_m$ and $T_m$ are the two parameters to be matched.

In cache networks with linear topology (i.e., tandem networks) the miss stream of a cache 
immediately provides the request stream to the following cache along the chain.
In tree-like topologies, instead, the request process arriving at a non-leaf cache is given by the 
superposition of the miss streams produced by children caches.
The inter-request distribution at non-leaf caches can be exactly characterized according to 
Theorem 4.1 in \cite{lawrence1973dependency}; however, we emphasize 
that the superposition of independent renewal process is not in general a renewal 
process~\cite{lawrence1973dependency}. Adapting the approach proposed in \cite{nain}, 
we approximately characterize the inter-request process at a non-leaf cache by an 
ON-OFF process whose {\em short} inter-request times are computed exploiting 
Theorem 4.1 in \cite{lawrence1973dependency}.

For example, for LRU and RANDOM, in the case of a cache having $K$ identical children 
whose miss streams are described by class \equaref{class1} (with parameters $\gamma_m$ and $T_m$),
we get:
\[
 F_{\text{short}}(t)=\left\{ \begin{array}{ll}    
\!\!\! 1\!-\!  \left(\frac{\gamma_m}{\gamma_m T_m +1 } \right)^{K-1} \!\!\! (T_m+ \frac{1}{\gamma_m} -t)^{K-1}  &  t\le T_m  \\
\!\!\! 1\!-\!  \left(\frac{1}{\gamma_m T_m +1 }\right)^{K-1} e^{-K \lambda (t-T_m)}  & t>T_m  \\
\end{array} \right.
\]
A similar expression (not reported here for the sake of brevity) is obtained for the 
class of inter-miss distribution \equaref{class2} adopted for the q-LRU policy.

\section{Evaluation of the cache hit probability}\label{sec:eval}
For completeness, we report here, for all caching policies
considered in this paper, the formulas to compute the hit probability
$p_{\text{hit}}(m)$ of an {\em arriving} request for object $m$, 
and the time-average probability $p_{\text{in}}(m)$ 
that object $m$ is found in the cache,
although these formulas have been already derived elsewhere \cite{che,nain,nostro_infocom}.
The overall hit probability $p_{\text{hit}}$ of a cache can be computed
by de-conditioning with respect to the content (Section \ref{subsec:decon}).

\subsection{LRU}
Under LRU 
we exploit the fact that object $m$ is found in the cache at time $t$ by an arriving request 
if and only if the previous request arrived in $[t-T_C, t)$:
$p_{\text{hit}}(m) = F_R(m,T_C)$.
The expression of $p_{\text{in}}(m)$ can be obtained exploiting the same argument,
but this time using the pdf $\hat{F}_R(T_C)$ of the {\em age} associated to object-$m$ 
inter-request time distribution: 
$p_{\text{in}}(m) = \hat{F}_R(m,T_C)$.

\subsection{q-LRU}
Under q-LRU, to compute $p_{\text{hit}}(m)$ we exploit the following reasoning:
an object $m$ is in the cache at time $t$ provided that: 
i) the last request arrived at $\tau\in[t-T_C,t)$ and ii) either 
at $\tau^-$ object $m$ was already in the cache, or its insertion was 
triggered by the request arriving at $\tau$  (with probability $q$).
We obtain: $p_{\text{hit}}(m) = F(m,T_C)[p_{\text{hit}}(m)+q(1-p_{\text{hit}}(m))]$.
The {\em age} distribution must be instead used to compute $p_{\text{in}}(m)$:
$p_{\text{in}}(m)=\hat{F}(m,T_C)[p_{\text{hit}}(m)+q(1-p_{\text{hit}}(m))]$. 
Once again, we emphasize that the argument above
{\em requires} the arrival process of requests to be stationary. As such, it can 
be hardly generalized to the case in which the request arrival process
is not stationary (like in SNM).

\subsection{RANDOM}
The decoupling principle of Che's approximation can be applied to the RANDOM caching policy 
by reinterpreting $T_C$ as the {\em random} sojourn time of a generic content in the cache,
whose distribution does not depend on the specific content. 
The eviction policy of RANDOM naturally leads to the choice of modeling $T_C$
as an exponentially distributed random variable. 
Under {\em renewal} traffic, the dynamics of each object $m$ in the cache 
can be described by a G/M/1/0 queuing model.
Indeed, the hit probability $p_{\text{hit}}(m)$ can be easily recognized to be equivalent to the
loss probability of a G/M/1/0 queue. Solving the Markov chain representing the 
number of customers in the system at arrival times, we get: 
$p_{\text{hit}}(m) = M_R(m,-1/\mathbb{E}[T_C])$,
where $M_R(m,\cdot)$ is the moment generating function of object-$m$ 
inter-request time.

Probability $p_{\text{in}}(m)$ can be obtained exploiting the fact that the 
dynamics of a G/M/1/0 system are described by a process that  
regenerates at each arrival. On such a process one can perform a 
standard cycle analysis \cite{nostro_infocom}, 
obtaining: $ p_{\text{in}}(m) = \lambda_m \,{\mathbb{E}[T_C]} \,(1-M_R(m,-1/{\mathbb{E}[T_C]})) $.

\subsection{2-LRU}
We assign index 1 and index 2 to the virtual and the physical cache, respectively. 
Let $T^i_C$ be the the eviction time of cache $i=1,2$. 
Cache 1 behaves exactly like a standard LRU cache, for which we can use previously derived expressions. 
An approximate analysis of cache 2 can be performed \cite{nostro_infocom}
by the following argument: object $m$ is found in cache 2 at time $t$ if and only if  
the last request arrived in $\tau\in[t-T^2_C,t)$ and either object $m$ was already in cache 2 at time $\tau^-$ 
or it was not in cache 2 at time $\tau^-$, but its ID was already stored in cache 1. 
Under the additional approximation that the states of cache 1 and cache 2 are independent 
at time $\tau^-$, we obtain:
\begin{eqnarray*}
p_{\text{hit}} (m) &\approx & F_R(m,T^2_C) [ p_{\text{hit}}(m)+ F_R(\lambda_m,T^1_C) (1-p_{\text{hit}}(m))] \\
p_{\text{in}} (m) &\approx &\hat{F}_R(m,T^2_C) [ p_{\text{hit}}(m)+ F_R(\lambda_m,T^1_C) (1-p_{\text{hit}}(m))] 
\end{eqnarray*}

\subsection{De-conditioning the hit probability}\label{subsec:decon}
For all considered cache policies, the final cache hit probability 
$p_{\text{hit}}$ is obtained de-conditioning with respect to $\lambda_m$ (i.e., $V_m$) 
\be
p_{\text{hit}}=  \mathbb{E}_{V}[p_{\text{hit} }(V_m)]=\int p_{\text{hit} }(v) \diff F_V(v) 
\label{phit} 
\ee
where we assume that request volumes $V_m$ of different contents are i.i.d.
Note that, similarly to the basic IRM case \cite{che},
$T_C$ is computed exploiting the fact that $C$ by construction equals the sum of the $p_{\text{in}}(m)$'s:
\[
C= \sum_m p_{\text{in}}(m)= M \cdot \mathbb{E}_{V}[p_{\text{in} }(V_m)]=M \int p_{\text{hit} }(v) \diff F_V(v) 
\]

\section{Numerical results}\label{sec:results}
We now present a selection of numerical results, having two goals in mind:
first, to prove the accuracy of the analytical approximations
developed in previous sections to obtain the hit probability
of individual and interconnected caches, under different cache policies 
and replication strategies. We will achieve these goals comparing
analytical predictions for the hit probability with simulation results
obtained from an ad-hoc, event-driven simulator fed by the same ON-OFF
traffic considered in the analysis. Second, we will exploit the model
to analyse more complex scenarios (too expensive to explore by simulations)    
and provide interesting insights into the impact of dynamic contents
on cache performance. 
 
\subsection{Single cache}\label{subsec:single}
We start considering the basic case of one cache fed by a
single-class ON-OFF traffic model. We assume that the average
number of requests ($V_m$) attracted by each content follows a Pareto
distribution:
\( f_V(v)=\beta V_{\min}^\beta/v^{1+\beta} \), for $v\ge
V_{\min}$\footnote{Recall that the second moment of the Pareto distribution is finite for
$\beta>2$.}.
The choice of a Pareto distribution for $V_m$ is justified by the following two facts:
first, previous work have already proved that the popularity of several types of contents
(e.g., movies, songs, user-generated videos), i.e., the long-term number of requests 
attracted by each content, is well described by the Zipf's law~\cite{zipf,Roberts1}; 
second, a Zipf-like distribution is obtained when a large number of individual
content request volumes are independently generated
according to a Pareto distribution.

For the experiments presented in this section, we fix the average
number of requests for each content to $\mathbb{E}[V]=10$, and the average OFF period duration 
$T_{OFF}= 9 T_{ON}$. Furthermore  we fix  the arrival rate of new contents  $\gamma=50,000$  and derive  from  \eqref{gamma}
the correspondent  catalogue size  (it turns out  $M=500,000 \cdot T_{ON}$). In our plots, error bars
correspond to $95\%$ confidence intervals derived from simulation.
 
Fig.~\ref{fig:figura1} shows the hit probability achieved 
by the LRU policy as function of the cache size, for different values of the average
ON period duration $T_{ON}$ (the absolute time unit is not important, let's assume  it corresponds to 1 day), and $\beta=2$. 
We observe an almost perfect match between simulation results (the vertical error-bars appear as
points) and the model predictions (the lines).  Observe, however, that we could not run simulations for the case $T_{ON}=300$ due to memory constraints.
As expected, cache performance is deeply impacted by the average life-span of contents
($L=T_{ON}$). Indeed, for a given cache size, the hit probability is
roughly inversely proportional to $T_{ON}$. This confirms that capturing 
temporal locality in the traffic is of paramount importance while
developing analytical models for cache performance. 

To investigate the impact of the content popularity distribution, i.e., of the number of
requests attracted by a content ($V_m$), Fig.~\ref{fig:figura2} shows the 
hit probability achieved by LRU while varying the value of the Pareto exponent 
$\beta$, and keeping fixed $\mathbb{E}[V]=10$. In this scenario $T_{ON}$ has been set to $7$ (days). We observe again a very good
match between analysis and simulation.
Also the distribution of content request volumes plays an important role on cache performance:
the hit probability increases when the popularity distribution has a heavier tail 
(i.e., as we decrease $\beta$). 
Note, however, that the impact on cache performance of the specific value of $\beta$ is rather limited
when $\beta>2$ (i.e. when the variance of the content request volumes is finite), which is
the most common case encountered in practice (e.g., YouTube videos).
This fact marks a significant difference with respect to the classical IRM model 
(more in general, when contents are not dynamic) where the impact of the power-law
exponent of content popularity is always very large over its entire domain~\cite{Roberts_mix}. 

Fig.~\ref{fig:figura3log} compares the performance of different caching policies, 
in the case of $T_{ON} = 7$,  $\beta = 2$. In particular, we consider
LRU,  q-LRU with $q=0.1$, RANDOM and 2-LRU. We observe again a good agreement
between analysis and simulation. We emphasize that, in the case of dynamic contents, 
an analytical estimation of the cache hit probability for policies different from LRU
is in general very hard to obtain. To the best of our knowledge,
we are the first to propose a viable approach to predict the performance
of q-LRU, RANDOM and 2-LRU in the presence of dynamic contents, 
with remarkable degree of accuracy, despite the long list of approximations.

As already observed by other authors in the case of renewal
traffic~\cite{amiciburini,nostro_infocom}, 2-LRU and q-LRU outperform LRU and RANDOM when the cache size is 
small, since these policies produce the desirable effect of filtering out a significant portion of unpopular 
contents, leading to a better exploitation of the limited cache space. Note, however,  
that 2-LRU provides significantly better performance than q-LRU, since its 
filtering action is more effective and selective.
As we increase the cache size, the presence of an insertion filter (especially for q-LRU) becomes at 
some point counter-productive, as demonstrated
by the fact that curves related to both LRU and RANDOM eventually cross 
both q-LRU and 2-LRU curves. We also observe that LRU provides slightly 
better performance than RANDOM, although the impact of the eviction policy is rather small 
over the entire range of cache sizes. Due to its simplicity, RANDOM turns out to be 
a viable alternative to LRU, especially for the implementation of caches in the network core.
Fig.~\ref{fig:figura9} reports the hit probability achieved by the above  caching policies 
under the Youtube traffic trace already used in Fig-\ref{fig:phit_vs_cachesize_fertility_flip_trace_04}.
Observe that the ranking among the considered policies 
is exactly the same as in Fig.~\ref{fig:figura3log}.  


\sidebyside{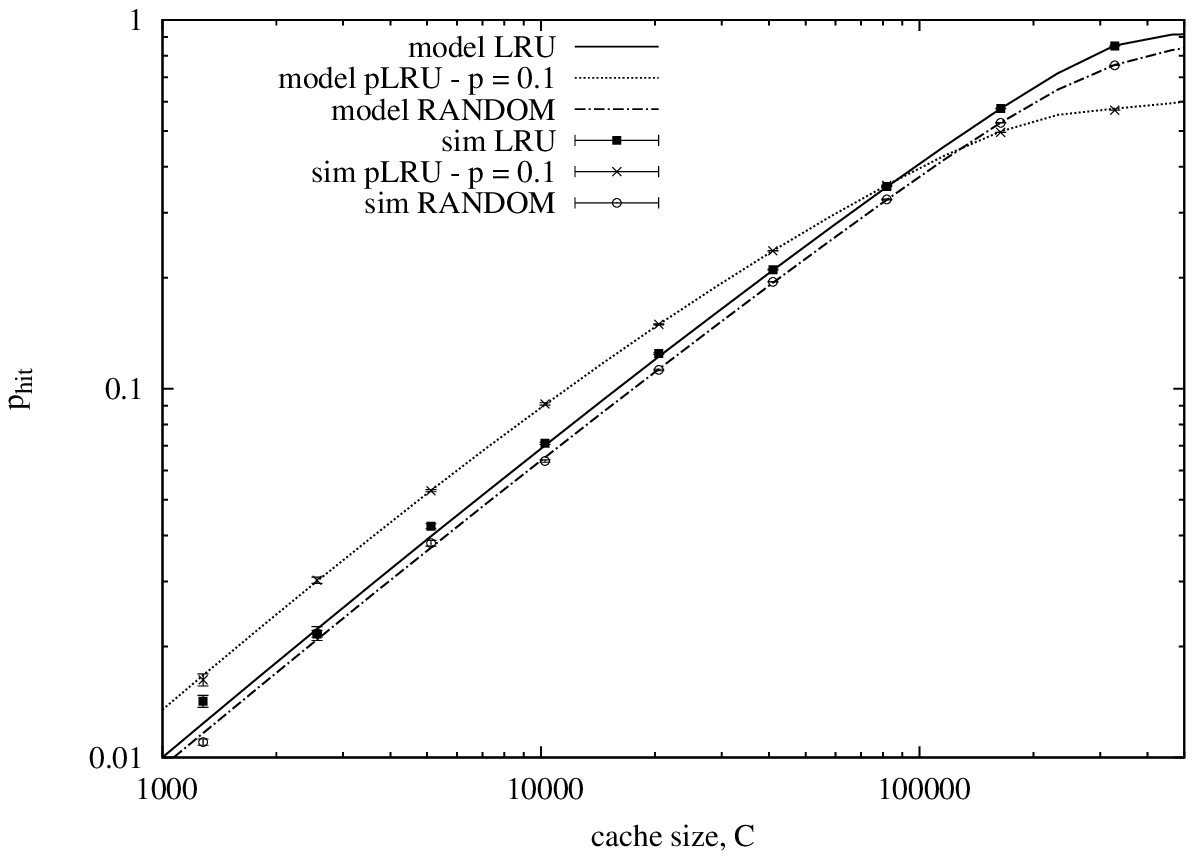}{Hit probability (in log scale) vs cache size, for different caching policies,
in the case of $T_{ON} = 7$,  $\beta = 2$.} {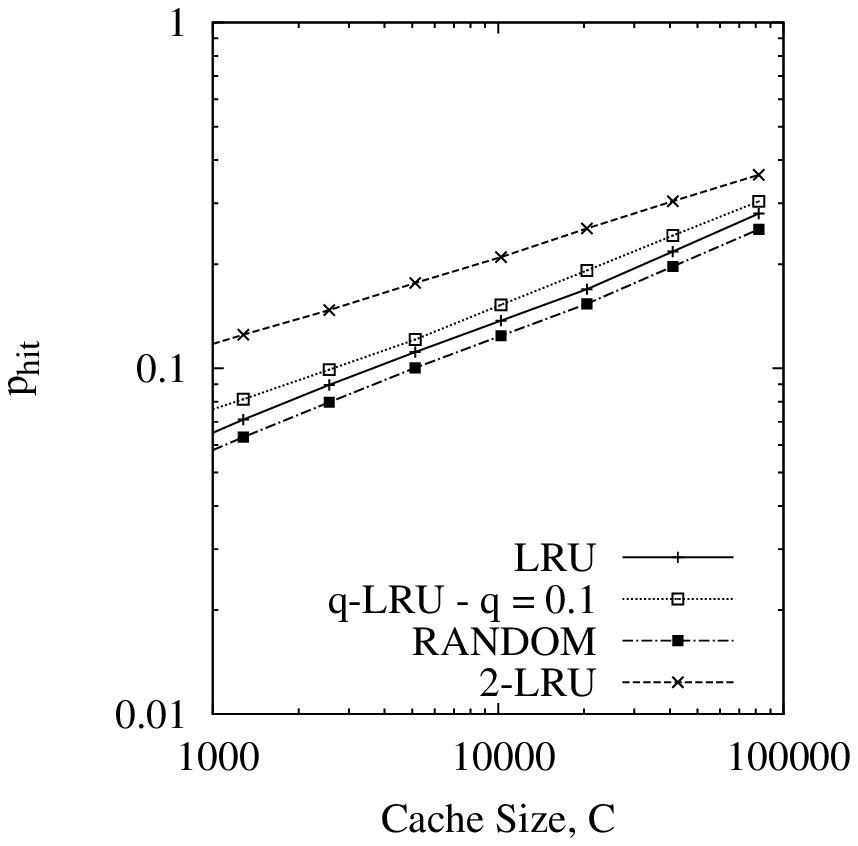}{Hit probability (in log scale) vs cache size, for different caching policies, under Youtube traffic trace.}{4.2}

\subsection{Cache networks}\label{subsec:binary} 
We now evaluate the accuracy of our model in cache networks.
In particular, we consider a tree-like topology of 15 caches (plus the repository above the root)
arranged as a binary tree with four layers. 
Also in this case we set the size of the content catalogue to $M = 10,000,000$, 
and we assume that the number of requests ($V_m$) attracted by a content 
at each of the 8 leaves follows a Pareto distribution with average $\mathbb{E}[V]=10$ and $\beta=2$. 
The average duration of the ON period is set to $T_{ON}= 7$ days (while $T_{OFF}=63$ days).
We consider two scenarios: 1) all caches in the tree have the same size;
2) the sum of cache sizes on each layer of the tree is the same (i.e., the size of a parent cache 
equals the sum of its children sizes).  

Fig. \ref{fig:figura4} reports the hit probability achieved by LRU, 
RANDOM and q-LRU (with $q=0.25$) in scenario 1.
We first observe that model predictions match very well simulations results also in the more
challenging case of a cache network.
Second, we observe that the gain achieved by q-LRU  with respect to LRU is even more
significant than in the case of a single cache (note that a filtering probability
$q=0.25$ obtains a gain similar to that of Fig.~\ref{fig:figura3log}, where
however we used $q=0.1$). 
Indeed, recall that assuming a q-LRU policy at each cache 
is equivalent to adopting the LCP replication strategy (leave-copy-probabilistically)
in an network of LRU caches. A probabilistic insertion policy
allows a better exploitation of the aggregate storage capacity of the system, by avoiding 
the simultaneous placement of an object in all caches along the path (note that, 
using $q=0.25$, we store on average only one copy along each route, given that the tree
has four layers).

Quite surprisingly, we observe that even the adoption of the RANDOM policy provides better performance
than LRU, in contrast to what we observed in the case of a single cache.
The superior performance of RANDOM with respect to LRU (assuming LCE replication)
was already shown in \cite{muscariellosig} for a tandem network, 
and it is confirmed here in the more general case of a tree-like network.
 
\sidebyside{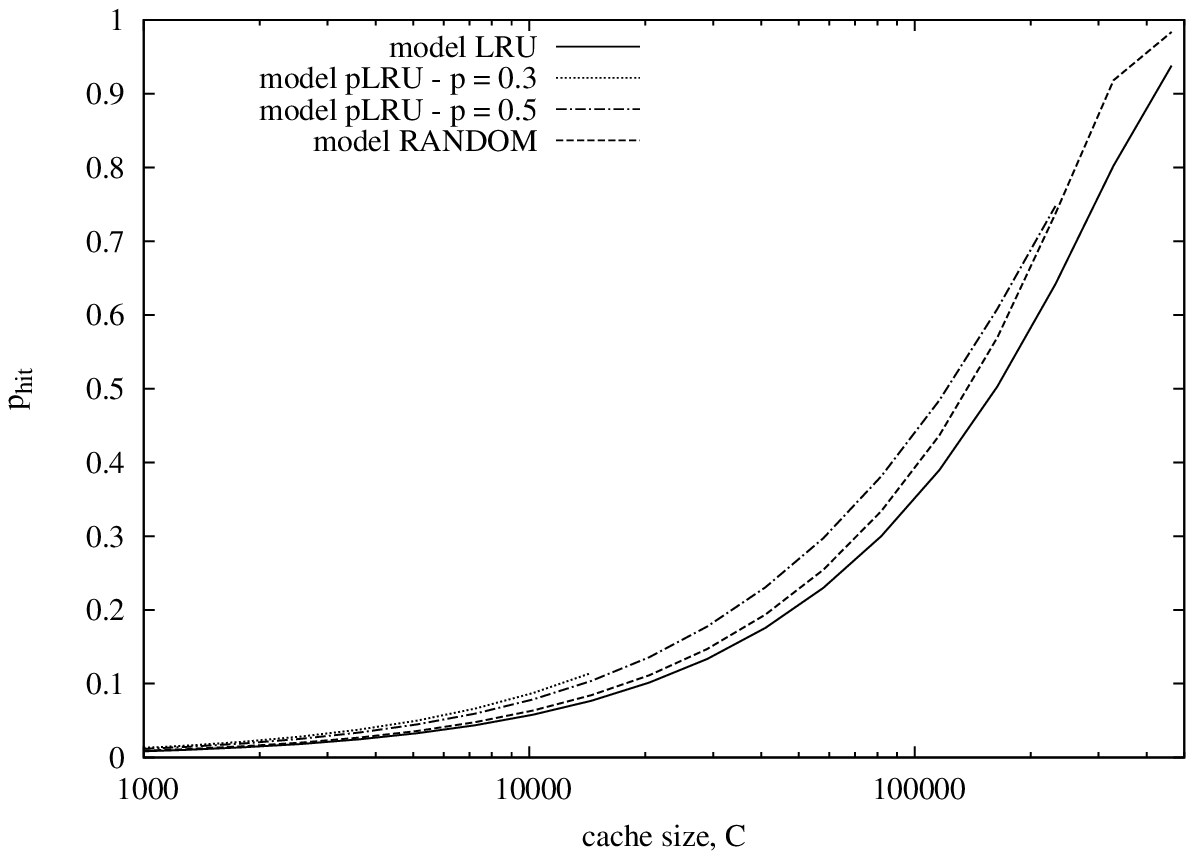}{Hit probability vs leaf cache size. Caches are all of the same size.}
{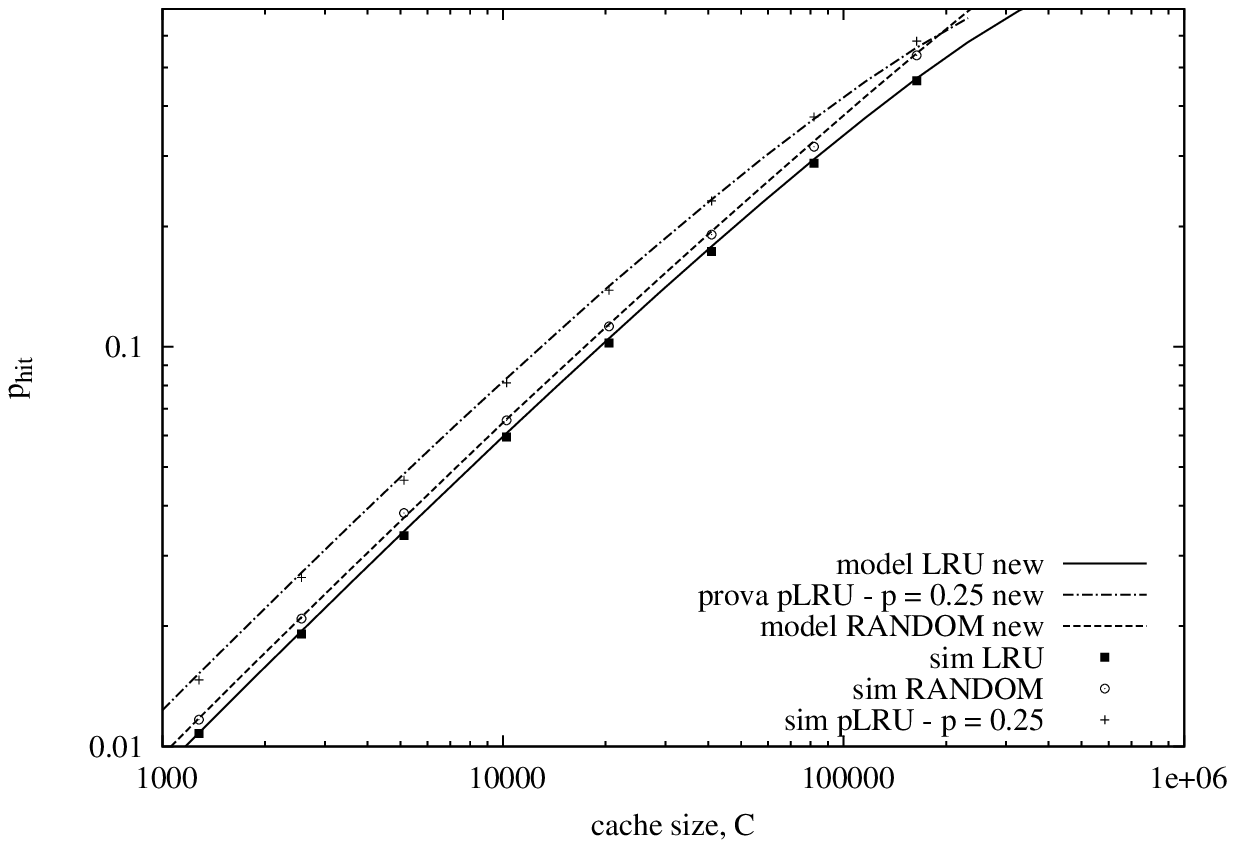}{Hit probability vs leaf cache size. The amount of storage at each layer of the tree is the same.}{4.2}

Fig. \ref{fig:figura5} complements previous analysis 
reporting the results obtained in scenario 2, where the size of a cache is set equal to the sum of the 
capacities of its children. Considerations analogous to those of scenario 1 can be drawn here.
As expected, for the same leaf cache size the overall hit probability
in scenario 2 is higher, thanks to the larger size of caches encountered going  
up along the tree.

\subsection{A realistic scenario}
Having validated the single-class model for both isolated and interconnected caches, we now 
consider the same binary-tree network examined in Section \ref{subsec:binary},
this time fed by a more realistic multi-class traffic,
showing how our approach can be effectively employed for system design and
optimization. We will only report analytical results here, since simulation results
were too expensive to obtain in this more complex scenario (this fact further
strengthens the usefulness of our methodology).
Our goal is to better understand the impact on cache
performance of a mixture of highly heterogeneous contents characterized
by different degrees of temporal locality. This is indeed 
the typical traffic observed in real networks \cite{TraversoCCR}.

In particular, we consider a mix of $6$ classes of contents, whose parameters,
listed in Table~\ref{tab:mix1}, have been chosen to reasonably represent 
the content heterogeneity produced by the popular YouTube platform, according
to measurements reported in \cite{TraversoCCR}.
Class~0 collects unpopular contents having request volumes smaller than 10.
Classes $1$--$5$ correspond to popular contents having different degree of
temporal locality, with average life-span ($L$) ranging from a few
days (Class~1) to several years (Class~5). 

\begin{table}[t!]
\centering
\resizebox{0.85\columnwidth}{!}{
\begin{centering}
\begin{tabular}{|c||c|c|c|c||c|c|c|}
\hline
Class &  ${L}$ (days)	& $E[{V}]$ & $V_{\text{max}}$	 & $\beta$ & Scenario 1 & Scenario 2 & Scenario 3\\
\hline
 0 	& 1000 	 &  1.6     & 10                       &  2.5          &  4 $\cdot 10^9$	 & 4$\cdot 10^9$ &   4$\cdot 10^9$  \cr
 1 	& 2      & 83.33    & $\infty$                 &  2.5          &  2.5$\cdot 10^6$  &   0        & 0  \\
 2 	& 7      & 75.00    &  $\infty$                &  2.5          &  3$\cdot 10^6$   &    3$\cdot 10^6$      & 0	    \cr
 3	& 30     & 66.66    & $\infty$                 &  2.5          &  3$\cdot 10^6$    &   3$\cdot 10^6$   &  3$\cdot 10^6$	    \cr
 4	& 100 	 & 50.0     &	 $\infty$              &  2.5          &  3.5$\cdot 10^6$  &  3.5$\cdot 10^6$     &  3.5$\cdot 10^6$\cr
 5	& 1000 	 & 50.0     & $\infty$                 &  2.5          & 15$\cdot 10^6$   &   15$\cdot 10^6$      &  15$\cdot 10^6$  \cr
\hline 
\end{tabular}
\end{centering}}
\vspace{1mm}
\caption{Content class parameters and their composition for each multi-class scenario.}
\label{tab:mix1}
\vspace*{-0.5cm}
\end{table}

In order to understand the impact of different traffic mixes, we
consider 3 traffic scenarios in which we vary the proportion of each
class of contents. This is equivalently obtained by varying the catalog size of contents 
belonging to the various classes,
as reported in the last $3$ columns of Table~\ref{tab:mix1}. 
Note that Class~1 is missing in both \textit{Scenario 2} and \textit{Scenario
3}, whereas Class~2 is missing only in \textit{Scenario 3}.
The presence or not of these two classes has been altered on purpose, since,
having the smallest value of content life-time $L$, they are expected 
to have the major impact on the overall hit probability (i.e., to be the more
\lq cacheable' classes of the mix).  

Fig.~\ref{fig:figura7} shows the performance of q-LRU (with $q = 0.25$) 
for the three considered scenarios, either in the case of caches all of the same size
(curves labelled \lq equal caches) or in the case of caches of size equal to the sum of their 
children (curved labelled \lq big caches'). 
We observe that {\em the presence of just a small fraction of highly
cacheable contents} (e.g., in \textit{Scenario 1}) {\em has a significant 
beneficial impact on the overall hit probability, especially with small
caches.}  Even for medium-size caches the gain is very significant: for
example, in the case of $C = 20000$, the hit probability  observed in  
\textit{Scenario 1} (around 0.1) is about twice the hit probability 
observed in \textit{Scenario 3}.



We now focus on \textit{Scenario 1} (where all classes are present), considering the 
case in which all caches have the same size.
This time, we assume that the system is able to restrict the access to the caches
only to contents belonging to a specific set of classes. Notice that this requires the
ability to classify objects' requests according to an a-priori knowledge of the 
popularity class they belong to.
This scenario is different from the one considered in Fig.~\ref{fig:figura7},
because this time requests for contents whose access into the cache is denied,  
deterministically produce a miss (whereas in the experiment in Fig. ~\ref{fig:figura7}
some classes simply where not present in the arrival stream of requests).
Now we are interested to see what happens when contents that are either unpopular (class \textit{0}) or
popular but long-lived (class \textit{5}) are not allowed to be cached.

\sidebyside{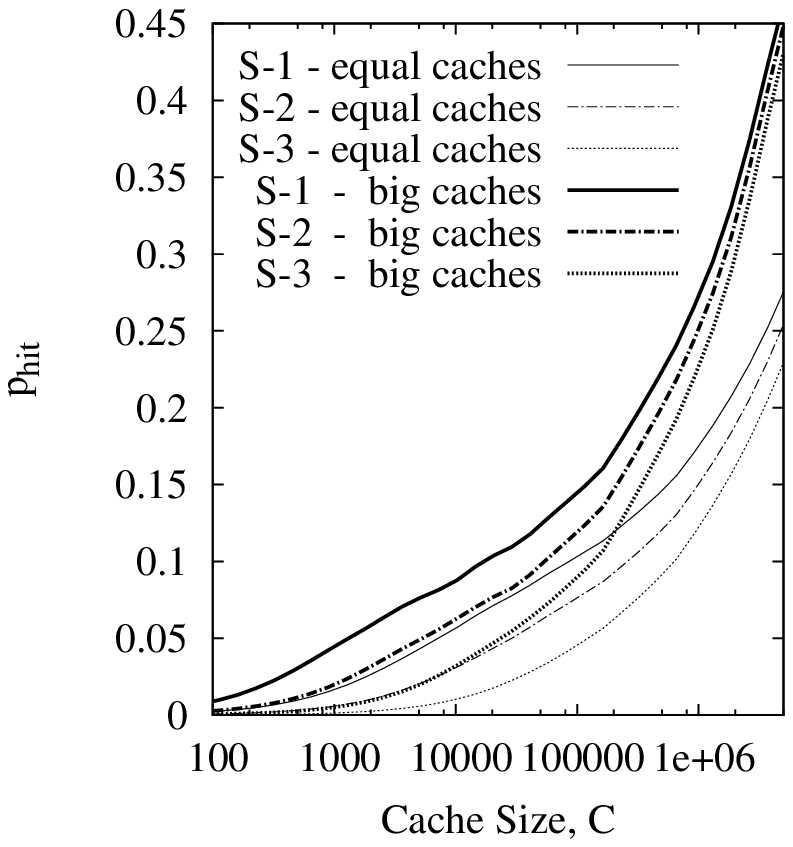}{Hit probability vs leaf cache  size for different traffic scenarios and cache
sizes, under q-LRU ($q = 0.25$).}{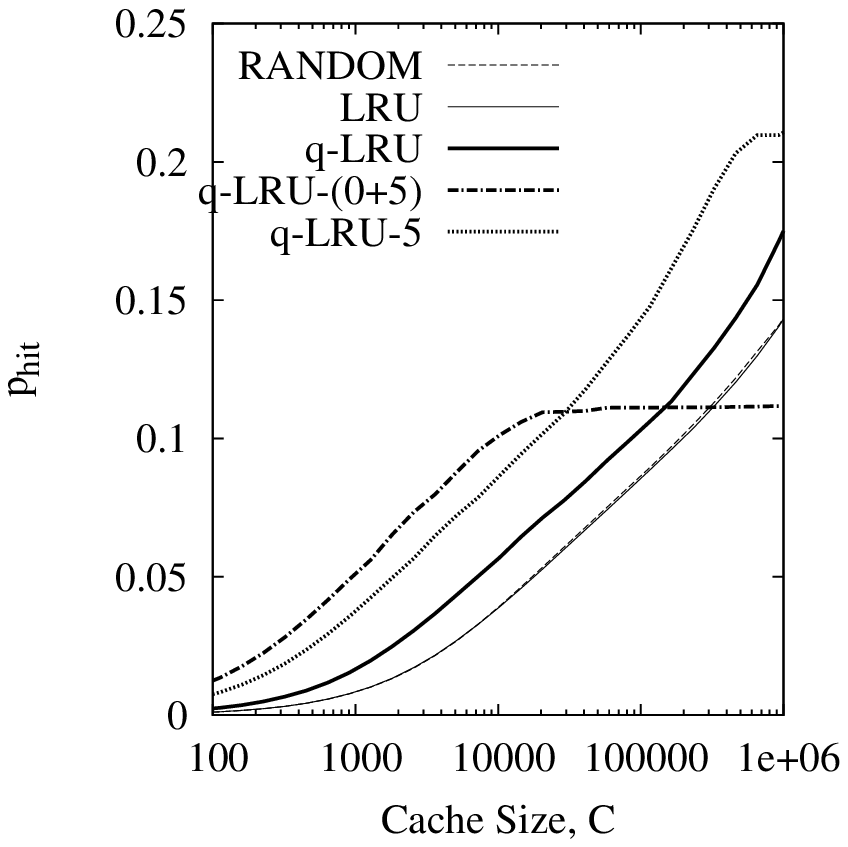}{Hit probability vs leaf cache size for 
different caching policies, with or without class filtering (in the case of q-LRU).}{4.2}
    
Fig~\ref{fig:figura8} compares the performance of LRU, RANDOM and  
q-LRU ($q = 0.25$) (without any class restriction) against the performance of 
q-LRU-0 and q-LRU-(0+5), where q-LRU does not cache contents of class \textit{0} and of 
both classes \textit{0} and \textit{5}, respectively. 

First, notice that q-LRU significantly outperforms both LRU and RANDOM (whose hit probability is nearly the same)
also in this more realistic scenario.
Second, we observe that, {\em when the cache
size is limited, a significant performance improvement is achieved by
filtering out contents that are either unpopular (class \textit{0}) or
popular but long-lived (class \textit{5}). } For example, the adoption
of q-LRU-(0+5) leads to a reduction of almost one order of magnitude (i.e., a factor of 10)
in the cache size that is needed to achieve $p_{\text{hit}}=0.1$, with
respect to q-LRU without access restrictions. 

As expected, filtering out contents when the
cache size increases must at some point become deleterious,
since filtered contents lead to a miss in the cache. This is confirmed by
the intersection between the curves in Fig~\ref{fig:figura8}.

The practical implementation of filters to detect unpopular/long lived
contents raises issues that go beyond the scope of this paper.

\section{Conclusions} \label{sec:concl}
We presented a general, accurate, and  computationally efficient 
approximate methodology for the analysis of  large distributed  systems
of interconnected  caches  under dynamic contents.  
Our methodology  can be successfully  applied to a  large  class  of caching strategies that 
includes  LRU, RANDOM  q-LRU and 2-LRU,
while maintaining the amenable property of representing request processes of individual 
contents with stationary processes.  
This is accomplished by modeling the requests arriving 
at different ingress caches with ``synchronized'' ON-OFF processes. 
We can then adapt and extend existing approaches based 
on the Che's approximation, inheriting all the nice properties of such approaches in terms 
of both accuracy and scalability.




 
\end{sloppypar}
\end{document}